\documentclass[letterpaper, 10pt, conference]{ieeeconf}      

\IEEEoverridecommandlockouts            
\makeatletter
\let\NAT@parse\undefined
\makeatother

\usepackage{graphicx}
\usepackage{color}
\usepackage{amsmath,amssymb,mathrsfs}  
\usepackage{amsthm}
\usepackage{amsfonts}
\usepackage[latin1]{inputenc}
\usepackage[english]{babel}
\usepackage{enumerate}
\usepackage[hidelinks]{hyperref} 
\usepackage{subfig}
\usepackage{comment} 
\usepackage[overload]{empheq}
\usepackage{bbm} 
\usepackage{tabstackengine}
\setstackgap{L}{3.5\normalbaselineskip}
\setstacktabbedgap{.01em}
\fixTABwidth{T}
\graphicspath{ {./figures/} }

\newcommand{\R}{\mathbb R}
\newcommand{\PL}{\mathcal P}
\newcommand{\E}{\mathcal E}
\newcommand{\T}{\mathcal T}
\newcommand{\F}{\mathcal{F}_\varepsilon}
\newcommand{\N}{\mathbb N}
\newcommand{\e}{e}
\newcommand{\V}{\mathcal V}
\newcommand{\K}{\mathcal K}
\newcommand{\I}{\mathcal I}

\newcommand{\rank}{\mathrm{rank}\,}

\newcommand{\blkdiag}{\mathrm{blkdiag}\,}

\newcommand{\C}{\mathbb C}

\newcommand{\bbm}[1]{\begin{bmatrix}#1\end{bmatrix}}
\DeclareMathAlphabet{\mathpzc}{OT1}{pzc}{m}{it}

\newtheorem{ass}{Assumption}

\newtheorem{alg}{Algorithm}

\newtheorem{prb}{Problem}
\newtheorem{prop}{Proposition}
\newtheorem{rem}{Remark}
\newenvironment{assumption}{\pushQED{\qed}\ass}{\popQED\endass}
\newenvironment{remark}{\pushQED{\qed}\rem}{\popQED\endrem}
\newenvironment{proposition}{\pushQED{\qed}\prop}{\popQED\endprop}

\allowdisplaybreaks[4]

\title{\LARGE \bf
Robust Hybrid Output Regulation for
Linear Systems\\ with Periodic Jumps:
the non-semiclassical case
}

\author{Giovanni de Carolis, Sergio Galeani, Mario Sassano
\thanks{G. de Carolis, S. Galeani, and M. Sassano are with the 
Dipartimento di Ingegneria Civile e Ingegneria Informatica, 
University of Rome Tor Vergata, 00133,  Italy, 
\texttt{[giovanni.de.carolis, sergio.galeani, mario.sassano]@uniroma2.it}. }}

\begin{document}
\let\pref\ref	
\renewcommand{\ref}[1]{\autoref{#1}}
\def\algautorefname{Algorithm}
\def\assautorefname{Assumption}
\def\defnautorefname{Definition}
\def\exmpautorefname{Example}
\def\factautorefname{Fact}
\def\figureautorefname{Fig.}
\def\lemautorefname{Lemma}
\def\probautorefname{Problem}
\def\propautorefname{Proposition}
\def\remautorefname{Remark}
\def\sectionautorefname{Section}
\def\subsectionautorefname{Section}
\def\thmautorefname{Theorem}
\def\corautorefname{Corollary}
\def\equationautorefname~#1\null{(#1)\null}
\allowdisplaybreaks[4]

\maketitle

\bstctlcite{IEEEexample:BSTcontrol} 

\begin{abstract}
This paper solves the robust hybrid output regulation problem for arbitrary uncertain hybrid MIMO linear systems 
with periodic jumps without the restrictive assumptions used in all previous works on the subject. 
A necessary condition for solving the problem is that 
the regulator must incorporate an internal model
of the flow zero-dynamics, which is typically affected 
by uncertainties and then unknown.
Hence, the proposed regulator consists of three units: 
a \emph{data-driven estimator} of the dynamics that are invisible from the regulated output during flows, 
a \textit{flow internal model} in charge of achieving regulation during flows, and 
a \textit{jump internal model} in charge of imposing a suitable reset of the state at each jump.
\end{abstract}

%
\section{Introduction}

The classic problem of output regulation, which includes as special cases the problems of reference tracking and disturbance rejection
when references/disturbances are deterministically generated by an exogenous system, is one of the key problems in control theory, 
perhaps second just to the stabilization problem. For this reason a lot of efforts have been devoted to solve this problem,
starting from the classic linear time invariant (LTI) setting considered in \cite{francis1976internal,davison1976robust,Wonham1985},
and then considering the nonlinear setting \cite{byrnes1997output,huang,pavlov2005uniform},
the use of adaptive or data driven mechanisms to estimate the exosystem's frequencies \cite{serrani_adaptiveIM,de2018data,2016_IJACSP},
the realization using an external (as opposed to internal) device still satisfying the internal model principle
\cite{Bodson1994,Messineo2009a,serrani2006rejection,2017_IFACWCa,2017_MEDb},
and contribution considering other structural features or constraints, like saturations on inputs and outputs \cite{Saberi} 
or overactuation \cite{galeani2015input,CDC_2011b,2018_ECCb,2018_CDC}.

In comparison with the above listed developments, 
the extension of output regulation theory to the case of hybrid systems has turned out to be more problematic.
Among the early works, contributions related to tracking in mechanical systems subject to impacts
(see \emph{e.g.} \cite{GMP_IJC_2008,morarescu2010trajectory,Forni,GMP_TAC12} and references therein)
have evidenced that, when considering hybrid systems whose jumps are state driven, 
even if the underlying flow and jump dynamics are linear, the resulting hybrid dynamics is strongly nonlinear,
and the derivation of sufficiently general and elegant results is very difficult.
Based on such evidence, the paper \cite{marconi2010note} introduced a much simpler hybrid output regulation problem
with time driven (and periodic) jumps, later studied in several papers including 
\cite{marconi2013internal,ZATTONI2017322,CDC_2012a,CDC_2012b,carnevale2016hybrid};
as is easy to see, if the underlying flow and jump dynamics are linear and the jumps are time driven,
linearity is preserved and the corresponding output regulation problem is amenable to 
an essentially linear analysis, parallel to classic results as in \cite{francis1976internal,davison1976robust};
such analysis is fundamental to gain highlights towards the understanding of the more complex, nonlinear case of hybrid output regulation with state driven jumps.
In particular, \cite{carnevale2016hybrid} provided a structural interpretation of the results in 
\cite{marconi2010note,marconi2013internal}, showing that the internal model unit needs to be able
to emulate not just the exosystem's dynamics, but also additional modes related to the flow zero dynamics
of a certain subsystem (the \emph{flow zero dynamics internal model principle}). 
The implication of such principle is that, since the zero dynamics is affected by plant uncertainties and 
must be considered unknown - thus making it impossible to directly replicate it in the internal model -, 
cannot be achieved by linear regulators unless special assumptions are made (as in \cite{marconi2013internal}, 
where the zero dynamics is implicitly required to be unaffected by parameter variations,
or as in \cite{2013_MEDc,carnevale2017robust}, where it is decoupled from the flow dynamics thanks to a physically motivated structure).
However, the solution of the robust hybrid output regulation problem in its generality,
\emph{i.e.} without the above restrictive assumptions, requires at least some form of adaptation of an 
otherwise linear regulator; such an approach, inspired by the non hybrid output regulation results in \cite{de2018data,2017_MEDb,2017_CDCc}, 
is pursued for the first time here, in the form of a data driven algorithm that tunes 
the regulator on the actual plant under control.

The paper is organized as follows: \ref{sec:prelim} establishes some preliminaries and the problem statement;
the design of the proposed regulator is detailed in \ref{sec:HR}; a numerical example showing the effectiveness 
of the proposed approach is presented in \ref{sec:example}; finally, some conclusions are provided in \ref{sec:example}.
An Appendix provides some details on the construction of a hybrid output feedback, observer-based stabilizer.
Although detailed proofs are omitted due to the (6 pages) space constraints of joint CDC/L-CSS submissions, 
motivating discussions appear along the paper to provide the line of reasoning which allows to prove 
the effectiveness of the proposed strategy.

\textbf{Notation}: For a matrix $M$, $ker(M)$ denotes its kernel, $im(M)$ denotes its image, and if $M$ is square, $\Lambda(M)$ denotes its spectrum (the set of its eigenvalues).
$\C_g$ represents the set of complex number with modulus less than one. The Kronecker product is denoted by $\otimes$.

\section{Preliminaries and Problem statement}\label{sec:prelim}
This paper focuses on the output regulation problem for a class of hybrid system, introduced in \cite{marconi2010note,marconi2013internal},
whose dynamics exhibit periodic jumps separated by a flow interval of known length $\tau_M > 0\,$.
As usual, two time variables $(t, k)$ are used, where $t$ measures the flow of time and $k$ counts the number of jumps.
In our scenario, admissible values of $(t, k)$ belong to a \textit{hybrid time domain} having the form:
\begin{align}\label{eq:hybrid_domain}
	\mathcal{T} = \{(t,k): t \in [t_k, t_{k+1}], k \in \N\}\,,\quad &t_k := k \tau_M\,.
\end{align}
Whenever the value of $(t,k)$ is clear from the context,
we consider the short-hand notations :
\begin{align*}
\dot{x}= \frac{d}{dt}x(t,k)\,,\quad &x^+=  x(t_k,k)\,.
\end{align*}
As discussed also in \cite{carnevale2017robust},  for a class of systems of the form
\begin{align}\label{eq:autonomous_system}
	\dot{x}_a= A_a x_a\,, \quad &x_a^+= E_a x_a\,, 
\end{align}
with time domain $\mathcal{T}\,$,
global exponential stability (GES) can be assessed by a simple test on the eigenvalues of the \textit{monodromy} matrix $\tilde{E}_a:= E_a e^{A_a\tau_M}$, without the need for a Lyapunov function; and, in particular, it is not necessary that $A_a$ be Hurwitz or $E_a$ be Schur. Moreover, since eigenvalues depend continuously  on the elements of $\tilde{E}_a\,$,
GES of \ref{eq:autonomous_system} implies GES also of the systems obtained considering small enough perturbations of $(E_a,A_a)\,$.\\
Consider the LTI hybrid plant $\PL$
\begin{subequations}\label{eq:linear_plant}
\begin{align}
	\dot{x} &= Ax+ Bu+ P w\,, \label{eq:linear_plant_flow}\\
	\e &= Cx + Q w \,, \label{eq:linear_plant_error}\\
	x^+ &= Ex\,,
\end{align}
\end{subequations}
where $x \in \R^{n}$, $u \in \R^m$ and $\e \in \R^p$ represent the state, the input and the output of $\mathcal{P}$ respectively;
$w\in \R^q$ acts as an exogenous input and represents the state of an exosystem $\mathcal{E}\,$
\begin{equation}\label{eq:exosystem}
\dot{w} = Sw\,,\qquad w^+ = Jw\,.
\end{equation}
Note that, the flows and jumps of \ref{eq:linear_plant} and \ref{eq:exosystem} are governed by the time domain $\T\,$, however the time arguments are neglected since they are univocally derived from $\T\,$.
The following assumption defines the class of models considered in this paper.
\begin{assumption}\label{ass:plant_exosystem_assumptions}{\rm
        The plant $\PL$ is over-actuated, namely $m > p\,$, and $\rank(B)=m$ and $\rank(C)=p\,$.
        Moreover, $\tilde{J}:= Je^{S\tau_M}$ is semi-simple and $\Lambda(\tilde{J}) \cap \C_g = \emptyset\,$.}
\end{assumption}
\begin{remark}{\rm
	The full-rank conditions on $B$ and $C$ are introduced to rule out trivialities.
    While $m\geq p$ is a well-know necessary requirement for output regulation in the presence of purely continuous-time systems,
    it has been shown in \cite{carnevale2016hybrid} that a prerequisite for output regulation robust to unstructured perturbation for the class of systems \ref{eq:linear_plant} governed by \ref{eq:hybrid_domain} consists in possessing strictly more inputs than outputs.
	Assuming $\tilde{J}$ to be semi-simple means that all its Jordan blocks have dimension one. 
	In addition, requiring that $\Lambda(\tilde{J}) \cap \C_g = \emptyset$ implies that no signal generated by \ref{eq:exosystem} for non-zero initial states asymptotically converges to zero.}
\end{remark}

\subsection{Problem Definition} 
The main objective of this paper consists in presenting a solution to the hybrid output regulation problem even if the underlying plant $\mathcal{P}$ in \ref{eq:linear_plant}  is described by an uncertain plant with a (known) nominal description $\mathcal{P}^0$.
In particular, denoting with $(A^0,B^0,P^0,C^0,Q^0,E^0)$ the state space characterization of $\mathcal{P}^0$ and let $(\Delta A,\Delta B,\Delta P,\Delta C,\Delta Q,\Delta E)$ be the possible perturbations affecting each matrix of $\mathcal{P}^0$,
namely the matrices in \ref{eq:linear_plant} are given by $A = A^0+ \Delta A\,$$\dots$,
we suppose that the plant $\PL$ belongs to a family $\F$ of admissible plants with $\varepsilon > 0$ such that $||\Delta||< \varepsilon\,$, $\Delta \in\{ \Delta A,\dots,\Delta E\}\,$.
\begin{prb}\label{prob:regulation_problem}
	Consider the nominal plant description $\PL^0$ of $\PL$ as in \ref{eq:linear_plant} and the exosystem $\E$ as in \ref{eq:exosystem}, and suppose that \ref{ass:plant_exosystem_assumptions} holds.
    Find, if any, a data-driven tuning algorithm for the error-feedback regulator 
    \begin{equation}\label{eq:hybrid_regulator}
    \begin{aligned}
    \dot{x}_c = A_c x_c+ B_c \e\,,\quad
    u = C_c x_c \,,\quad
    x_c^+ = E_c x_c\,,
    \end{aligned}
    \end{equation}
    ensuring, for some $\varepsilon>0\,$, it holds that for any $\PL\in \F$:
    	\begin{itemize}
        \item (GES) the interconnected system \ref{eq:linear_plant} , \ref{eq:hybrid_regulator}  with $w \equiv 0\,$ is globally exponentially stable;
        \item (OR) $\displaystyle{\lim_{t+k\to \infty}} \e(t,k)=0$ for all initial states of the interconnected system \ref{eq:linear_plant}, \ref{eq:exosystem}, \ref{eq:hybrid_regulator}. \hfill $\circ$
    \end{itemize}
\end{prb}
\begin{remark}{\rm
        A similar formulation of Problem \ref{prob:regulation_problem} (without the request for a tuning algorithm) is given in \cite{carnevale2017robust},
        where, however, it is assumed that the nominal matrices $A^0\,,B^0\,,\dots\,$, and the perturbations $\Delta A\,,\Delta B\,,\dots\,$, possess a specially structured partition such that the flow  zero-dynamics internal model principle discussed in \cite{carnevale2016hybrid} is trivially satisfied.
        This structural assumption, namely of dealing only systems in \textit{semiclassical form} (see \cite{carnevale2016hybrid}) has deep implications on the structure of the internal model units:
        robust output regulation can be achieved by incorporating only a copy of the modes of $S$ and $\tilde{J}\,$, and then there is no need for a tuning algorithm.
        In this paper we remove such structural assumption and solve the problem for general non-semiclassical systems.}
\end{remark}

\subsection{A Useful Decomposition}\label{sec:decomp}
As introduced in \cite{galeani2015input} and further stressed in \cite{carnevale2016hybrid}, there exists a coordinate change in the input, state and output spaces that, combined with a preliminary state feedback, is such to decompose the plant \ref{eq:linear_plant} in two subsystems which evolve separately during flows and are coupled at jumps.
The change of coordinates mentioned above is obtained by exploiting some geometrical concepts presented in \cite{galeani2015input} and recalled here for clarity.\\
Define $\V^\star \subset \R^n$ as the subspace of states for which there exists an input function such that the output of \ref{eq:linear_plant_flow}, \ref{eq:linear_plant_error} 
remains identically zero for all times,
and  define $\mathcal{R}^\star \subset \R^n$ as the subspace of states for which there exists an input function 
steering the state of \ref{eq:linear_plant_flow}, \ref{eq:linear_plant_error} to zero in finite time while keeping the output identically zero. 
Let $\rho := \text{dim} (\mathcal{R}^\star)$ and $\nu:= \text{dim} (\mathcal{V}^\star )\,$, where $\nu\geq \rho$ since $\mathcal{R}^\star\subset \mathcal{V}^\star\,$.
Choose $T\in\R^{n\times n}$ such that its first $\rho$ columns span $\mathcal{R}^\star$ and its first $\nu$ columns span $\mathcal{V}^\star$ whereas its last $n-\nu$ columns define a basis for a subspace $\mathcal{Z}$ such that $\mathcal{Z}\oplus\mathcal{V}^\star =\R^n\,$.
Choose $G = \bbm{G_1 &G_2}\in \R^{m \times m}$ as an invertible matrix such that $im(G_1) = B^{-1} \mathcal{R}^\star$ and choose $F^\star_{\mathcal{V}}$ to satisfy $(A+BF^\star_{\mathcal{V}}){\mathcal{V}^\star} \subset {\mathcal{V}^\star}\,$.
Then, applying the coordinate change 
\begin{equation}\label{eq:change_coordinates}
	z = T^{-1} x\,,
\end{equation}
and the regular feedback transformation
\begin{equation}\label{eq:feedback_transformation}
\begin{aligned}
u&=G(G^{-1}F^\star_{\mathcal{V}} Tz +  \bar{u})= G(\bar{F}^\star_{\mathcal{V}} z +  \bar{u}) \\
&= G \bar{v}\,,
\end{aligned}
\end{equation} 
where $\bar{u}=\bar{v}-\bar{F}^\star_{\mathcal{V}} z\,$, to the plant $\PL$,
the dynamics in the transformed coordinates are described by
\begin{subequations}\label{eq:linear_plant_new_coordinates}
	\begin{align}
	\dot{z} &= \bar{A}^Fz+ \bar{B}\bar{u}+ \bar{P} w \,,\label{eq:linear_plant_new_coordinates_a}\\
	\e &= \bar{C}z + Q w \,,\label{eq:linear_plant_new_coordinates_b}\\
	z^+ &= \bar{E}z\,,
	\end{align}
\end{subequations}
where $ \bar{A}^F = T^{-1}(A+BF^\star_{\mathcal{V}})T =\bar{A}+\bar{B}\bar{F}^\star_{\mathcal{V}}\,$, $ \bar{B} = T^{-1}B G\,$, $ \bar{P} = T^{-1}P\,$, $\bar{C} = CT\,$ and $ \bar{E} = T^{-1}ET\,$ have the form
\begin{equation*}\label{eq:partitioned_form}
	\begin{aligned}
		\bar{A}^F &=\bbm{\bar{A}^F_{11} &\bar{A}^F_{12} &\bar{A}^F_{13}\\ 0&\bar{A}^F_{22} &\bar{A}^F_{23}\\ 0 &0 &\bar{A}^F_{33}},
		\bar{B} =\bbm{\bar{B}_{11} &\bar{B}_{12}\\ 0 &\bar{B}_{22}\\ 0 &\bar{B}_{32}} = \bbm{\bar{B}_1 & \bar{B}_2},\\
		\bar{P} &=\bbm{\bar{P}_{1}\\ \bar{P}_{2}\\\bar{P}_{3}},	
	    \bar{E} =\bbm{ \bar{E}_{11} & \bar{E}_{12} & \bar{E}_{13}\\  \bar{E}_{21} & \bar{E}_{22} & \bar{E}_{23}\\  \bar{E}_{31} & \bar{E}_{32} & \bar{E}_{33}},
		\bar{C} =\bbm{0 &0 &\bar{C}_3}\,.
	\end{aligned}
\end{equation*}
\begin{remark}\label{rem:matrices_new_coordinates}{\rm
 By definition the pair $(\bar{A}_{11}^F, \bar{B}_{11})$	is controllable, hence the spectrum of $\bar{A}_{11}^F$ can be arbitrary assigned by an appropriate selection of $F^\star_{\mathcal{V}}$ as described in \cite{trentelman2002control}. Conversely, $\Lambda(\bar{A}_{22}^F)$ cannot be assigned by $F^\star_{\mathcal{V}}$ and it coincides with the set of invariant zeros of the plant \ref{eq:linear_plant}.}
\end{remark}

\subsection{Structural Conditions for Output Regulation}
The solvability of Problem \ref{prob:regulation_problem} for a plant possessing the structure of \ref{eq:linear_plant_new_coordinates} is discussed in \cite{carnevale2017robust}.
In particular, when only error measurements are available, the \textit{output regulation problem }is solvable for a given family $\F$ if and only if, $\forall \PL \in \F\,$:
\begin{itemize}
	\item[$a_h)$] The plant \ref{eq:linear_plant_new_coordinates} is stabilizable and detectable, namely
	\begin{subequations}\label{eq:structural_conditions}
	\begin{align}
		\rank(R_H(s))=n \qquad&\forall s \in \Lambda(\bar{E}e^{\bar{A}^F\tau_M}) \setminus \C_g\,,\label{eq:PBH_test}\\
		\rank(O_H(s))=n \qquad&\forall s \in \Lambda(\bar{E}e^{\bar{A}^F\tau_M}) \setminus \C_g\,,\label{eq:PBH_test_b}
	\end{align}
	where
	\begin{align*}
	R_H(s)&:=\bbm{\bar{E}e^{\bar{A}^F \tau_M}- sI &\mathcal{R}(\bar{A}^F,\bar{B})}\,,\\
	O_H(s)&:=\bbm{\bar{E}e^{\bar{A}^F \tau_M}- sI &\mathcal{O}(\bar{A}^F,\bar{C})}\,,\\
	\end{align*}
    with  $\mathcal{R}(\bar{A}^F,\bar{B}) := [\bar{B}\; \bar{A}^F\bar{B}\; \cdots (\bar{A}^F)^{n-1}\bar{B}]$ and $\mathcal{O}(\bar{A}^F,\bar{C})=[\bar{C}'\; (\bar{A}^F)'\bar{C}'\; \cdots ((\bar{A}^F)^{n-1})'\bar{C}']'\,$;
	\item[$b_h)$] The following non-resonance conditions hold:
	\begin{align}\label{eq:non-resonance_condition}
	\rank(P_{F,3}(s))&=n_3+p \qquad&\forall s \in \Lambda(S)\,,\\
	\rank(P_H(s))&=n \qquad&\forall s \in \Lambda(\tilde{J})\,,
	\end{align}
	\end{subequations}
	where
	\begin{align*}
	P_{F,3}(s)&:=\bbm{\bar{A}^F_{33}-sI &\bar{B}_{32}\\\bar{C}_3 &0}\,,\\
	P_{H}(s)&:=\bbm{\bar{E}_{11}&\bar{E}_{12} &\bar{E}_{11}\\
								\bar{E}_{21}&\bar{E}_{22} &\bar{E}_{21}\\
								\bar{E}_{31}&\bar{E}_{32}&\bar{E}_{31}}e^{\tilde{A}\tau_M} - s\bbm{I &0 &0\\0 &I &0\\ 0&0 &0}\,,
	\end{align*}
	with $\tilde{A}:=\blkdiag\tiny\left\lbrace\bbm{\bar{A}^F_{11}&\bar{A}^F_{12} \\ 0 &\bar{A}^F_{22}}\,,0\right\rbrace$
	and $n_3$ is the dimension of $\bar{A}^F_{33}\,$.
\end{itemize} 
It can be shown that the following property holds.
\begin{proposition}{\rm
    Suppose that conditions \ref{eq:structural_conditions} hold for $\PL^0$. Then there exists $\bar{\varepsilon}>0$ such that any $\PL\in\F$ with $\varepsilon\in(0,\bar{\varepsilon})$ satisfies \ref{eq:structural_conditions}.}
\end{proposition}
As a consequence, the following assumption can be considered towards the solution of Problem \ref{prob:regulation_problem}.
\begin{assumption}\label{ass:structural_Conditions}{\rm
	The nominal plant $\PL^0\,$ satisfies \ref{eq:structural_conditions}.}
\end{assumption}
\begin{remark}{\rm
    The conditions \ref{eq:structural_conditions} are formulated for a system of the form proposed in \ref{eq:linear_plant_new_coordinates} for simplicity. Nonetheless, it can be easily shown that $\PL^0\,$ in \ref{eq:linear_plant} satisfies such conditions if and only if the plant transformed via \ref{eq:change_coordinates}, \ref{eq:feedback_transformation}
    satisfies \ref{eq:structural_conditions}.}
\end{remark}
\section{Hybrid Regulator}\label{sec:HR}
\begin{figure}[t]
    \centering
    \includegraphics[width=0.7\columnwidth]{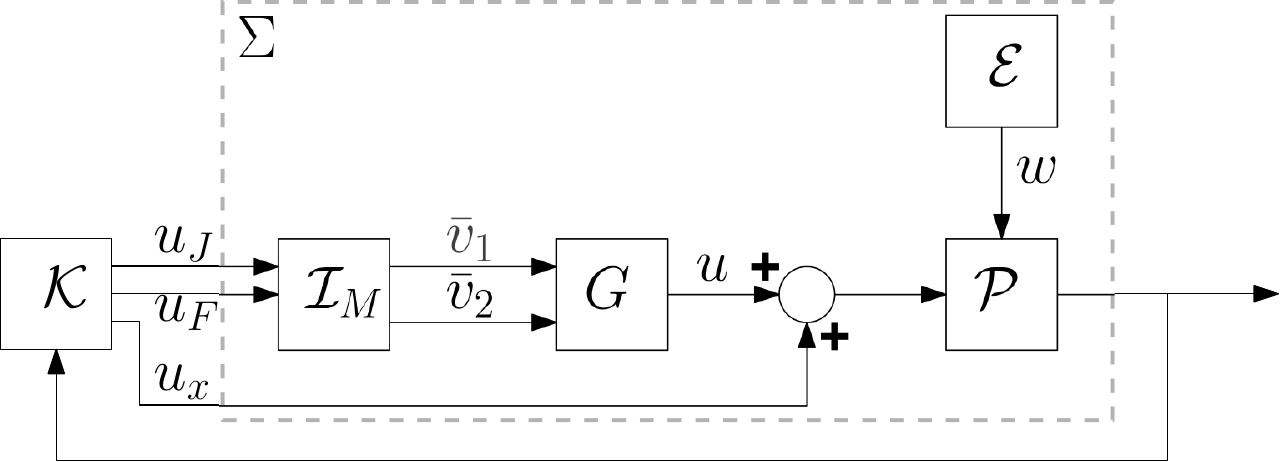}
    \caption{The internal model based regulator.}
    \label{fig:regulator}
\end{figure}
As pointed out in \cite{carnevale2017robust},
under Assumptions \pref{ass:plant_exosystem_assumptions} and \pref{ass:structural_Conditions},
a regulator of the form of \ref{eq:hybrid_regulator} can be designed, based only on knowledge
of the exosystem and of the nominal description $\PL^0\,$, to guarantee (OR) in Problem \ref{prob:regulation_problem} for all plants $\PL \in \F$ having the form described in \ref{eq:linear_plant_new_coordinates} and satisfying \ref{eq:structural_conditions} for which the system in closed-loop with a controller designed on the knowledge of $\PL^0$ preserves the global exponential stability.\\
Such regulator consists of the so called \textit{heart of the hybrid regulator}, whose role is to solve a classic purely continuous-time output regulation problem associated to the system $(\bar{A}^F_{33},\bar{B}_{32},\bar{C}_{3},0)$ with exogenous input $w$ entering through matrices $\bar{P_3}$ and $Q\,$, and of a second internal model unit that takes care of an auxiliary regulation problem at jumps.
In particular, the former unit must contain a copy of the modes of $S$ while the latter must incorporate a copy of those of $\tilde{J}\,$.
\begin{remark}{\rm
    Since in \cite{carnevale2016hybrid} the plant is assumed to belong to the class of systems described by \ref{eq:linear_plant_new_coordinates} that satisfy \ref{eq:structural_conditions},
    hence it is not required to include the term $-\bar{F}^\star_{\mathcal{V}} z$ in \ref{eq:feedback_transformation},
    the construction of the internal model units can be carried out only relying on information about the exosystem that is assumed precisely known.
    Here, instead, due to the generic presence of the term $-\bar{F}^\star_{\mathcal{V}} z$ in \ref{eq:feedback_transformation},
    a solution to Problem~\ref{prob:regulation_problem} can be obtained only having a precise knowledge of the exosystem $\E$ and of the dynamics of $\bar{A}^F_{11}$ and $\bar{A}^F_{22}\,$.
    However, while the former is assumed given, the latter are in general affected by uncertainty.
    Therefore, an estimation procedure becomes necessary.}
\end{remark}
\subsection{Estimation of $\bar{A}^F_{11}$, $\bar{A}^F_{22}$ and $G$}\label{app:estimator}
Consider the uncertain system \ref{eq:linear_plant} and suppose that
the pair $(A,C)$ is observable.
By defining a zero-order-hold (ZOH) with sampling time $\tau\,$,
the following discrete-time description
\begin{equation}\label{app:simple_discrete_description1}
    \begin{aligned}
        \eta_{[i+1]} = A_D\eta_{[i]}+ B_Du_{[i]}\,,\;
        \eta^+_{[n]} = E\eta_{[n]}\,,
        y_{[i]} = C\eta_{[i]}  \,,\;
    \end{aligned}
\end{equation}
with $i\in \mathbb{N}$, can be obtained
where
\begin{align}\label{eq:app_descrete_relation1}
A_{D} = e^{A\tau},\; B_{D} = \left(\int_{0}^{\tau}e^{A\theta} d\theta\right) B\,.
\end{align}
Let $[1,a_{n-1},\dots,a_{0}]$ be the coefficients of the characteristic polynomial of $A_D$, then the flow dynamics of \ref{app:simple_discrete_description1} are immersed in those of a system described by
\begin{equation}\label{eq:app_characterization1}
\begin{aligned}
\eta_{O[i+1]} &= A_O\eta_{O[i]}+ B_Ou_{[i]}\,,\\
y_{[i]} &= C_O\eta_{O[i]}  \,,\\
\end{aligned}
\end{equation}
where 
\begin{align*}
A_O &:= I_{p} \otimes A_{O0} \in \R^{np\times np}\,,\\
C_O &:= I_{p} \otimes C_{O0} \in \R^{p\times np}\,,\\
B_O&:=\bbm{B_{O1}'\dots,B_{Onp}'}'\in \R^{np\times m}\,,
\end{align*}
and 
\begin{align*}
A_{O0}&= \bbm{0 &1 &0 &\dots &0\\ \vdots &\ddots &\ddots &\ddots &\vdots \\0 &\dots &0 &1 &0\\ \text{-}a_0 &\dots &\text{-}a_{n-3}&\text{-}a_{n-2}&\text{-}a_{n-1}}\,,\\
C_{O0} &=\bbm{\phantom{+}1&\phantom{+}0&\phantom{+}\dots&\phantom{++}0 &\phantom{++}0\phantom{+}}\,.
\end{align*}
A straightforward implication of the dynamics \ref{app:simple_discrete_description1} is that
\begin{align}\label{eq:app_new_relation}
Y_{[i]}= \eta_{O[i]} + \tilde{D} U_{[i]}\,,
\end{align}
where
\begin{align*}
\tilde{D}=\bbm{0 &\cdots &0 &0\\
    B_{O1} &\ddots  &\vdots &\vdots\\\vdots &\ddots  &0&0\\ B_{Onp-1} &\dots &B_{O1}&0},
\end{align*}
and $Y_{[i]} = [y_{[i]}',\cdots,y_{[i+n-1]}']' $, $U_{[i]} = [u_{[i]}',\dots,u_{[i+n-1]}']'\,$ are a collection of $n$ consecutive measurements of the inputs and the outputs.
At the same time, defining $a := [a_0,\dots,a_{n-1}]'$, provided the input is identically
equal to zero,
equation \ref{app:simple_discrete_description1} and \ref{eq:app_new_relation} can be combined to obtain that
\begin{align}\label{eq:app_solution}
y_{[n+i]}= (-a'\otimes I_p) Y_{[i]} = -\bar{Y}_{[i]}a\,, 
\end{align}
where $\bar{Y}_{[i]} = [y_{[i]},\cdots,y_{[i+n-1]}]\,$.
Therefore, imposing $u_{[h]} = 0$ for $h = 0,\dots,2n-1$,
equation \ref{eq:app_solution} can be recursively exploited to conclude that
\begin{align}\label{eq:app_ToInverse}
Y_{[n]}=-\mathcal{Y} a\,, 
\end{align}
with $\mathcal{Y}= -\bbm{\bar{Y}_{[0]}',\dots,\bar{Y}_{[n-1]}'}'\,$.
\begin{figure}[t]
    \centering
    \includegraphics[width=0.45\columnwidth]{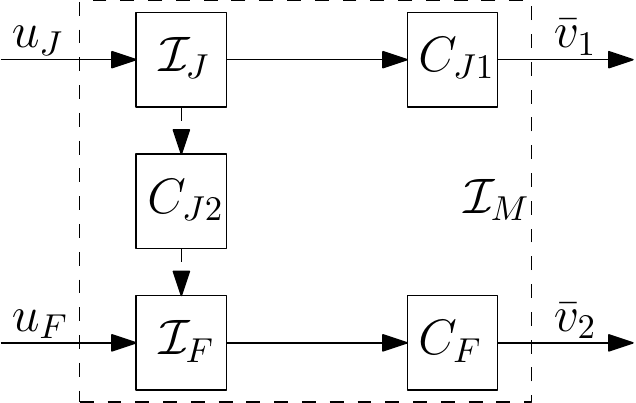}
    \caption{Structure of the internal model $\mathcal{I}_M$.}
    \label{fig:internal_model}
\end{figure}
\begin{proposition}{\rm
        If the matrix $\mathcal{Y}$ is full row rank, 
        then 
        $$a=-\mathcal{Y}^{\dagger} Y_{[n]}\,,$$
        with $\mathcal{Y} ^{\dagger} $ denoting the Moore-Penrose pseudo inverse of $\mathcal{Y}\,$.}
\end{proposition}
Knowing $A_O$, $C_O$ and $\eta_{O[2n-1]}=A_O^nY_{[n-1]}$, each column of the matrix $B_O$ can be calculated imposing $m(np+1)$ consecutive samples of the input equal to $I_m \otimes \mathcal{U}_0$, with $\mathcal{U}_0= [1,0,\dots,0]'\in\R^{np+1}\,$ and exploiting the fact that
\begin{align*}
\bbm{Y_{[i]}\\y_{[i+n]}}  =
\bbm{I_{np}\\C_OA_O^{n}} \eta_{O[i]}+
\bbm{0 &\cdots &0\\
    B_{O1} &\ddots  &\vdots \\\vdots &\ddots  &0\\ B_{Onp} &\dots &B_{O1}} U_{[i]}\,.
\end{align*}
Once a discrete time estimation of the considered system \ref{eq:linear_plant} has been obtained,
the corresponding continuous time estimation can be obtained inverting relations \ref{eq:app_descrete_relation1}.
Through the procedure described in \ref{sec:decomp},
a description of the continuous time evolution of \ref{eq:linear_plant} as in \ref{eq:linear_plant_new_coordinates_a}, \ref{eq:linear_plant_new_coordinates_b} is easily obtained and the matrices $\bar{A}^F_{11}$, $\bar{A}^F_{22}$ and $G$ can be calculated.
\subsection{Regulator design}
As shown in \ref{fig:regulator}, the proposed regulator is composed of two main dynamical blocks: an ``internal model'' $\mathcal{I}_M$ and a dynamic stabilizer $\mathcal{K}\,$. 
The internal model is composed by a \textit{jump internal model} $\I_{J}$ and a \textit{flow internal model} $\I_{F}$ interconnected as in \ref{fig:internal_model}.
In turn, the dynamic stabilizer $\mathcal{K}$ is made-up by a discrete-time controller $\K_D$ with a discrete-time observer $\mathcal{O}_D$, both designed on the same step size time, whose task consists in stabilizing the system $\Sigma$ in \ref{fig:regulator}
\subsubsection{The Internal Model $\I_M$}\label{subsec:internal_model}
The jump internal model $\I_J$ has to be designed to provide at each period the correct initialization of $\PL$ and $\I_{F}$ such to ensure $\e(t,k)=0$ for all $(t,k)$ with $t\in[t_k,\,t_{k+1}]\,$.
To that end, the jump internal model must contain $m_1+n_F$ independent copies of the dynamics of the exosystem,
where $m_1$ is the size of the plant input that acts through the first column block of $\bar{B}$ in \ref{eq:linear_plant_new_coordinates},
while $n_F$ represents the dimensions of $\I_F$.
\begin{alg}\label{alg: flow_internal_model}
    \textbf{Design of} $\I_F$ (see \ref{fig:internal_model})\\
    Assume to have an estimation $\hat{A}^F_{11}$, $\hat{A}^F_{22}$ of the dynamics of $\bar{A}^F_{11}$, $\bar{A}^F_{22}$ as proposed in \ref{app:estimator}.
    Let $\mu_h(s)$ be the minimal polynomial of $\blkdiag(\hat{A}^F_{11}$, $\hat{A}^F_{22},S)\,$, and define $n_h:= deg(\mu_h(s))\,$.
    Let $A_{F0}\in\R^{n_h \times n_h}$ be the lower companion matrix with $det(sI-A_{F0}) = \mu_h(s)$ and $C_{F0} = [1\, 0 \,\dots \,0] \in \R^{1 \times n_h}\,$.
    Define $\I_{F}$ according to
    \begin{equation}\label{eq:flow_internal_model}
    \begin{aligned}
    \dot{x}_F = A_F x_F + u_F\,,\;\,x^+_F = C_{J2} x_J\,,\;\, y_F = C_F x_F\,,
    \end{aligned}
    \end{equation}
    where $A_F= I_p \otimes A_{F0}\,$, $C_F= I_p \otimes C_{F0}\,$, $x_F\in\R^{n_F}\,$ and $n_F=pn_h\,$.
\end{alg}
\begin{alg}\label{alg: jump_internal_model}
	\textbf{Design of} $\I_J$ (see \ref{fig:internal_model})\\
	Let $q$ be the dimension of $w$ and $C_{J0} = [0\, \dots \,0\, 1] \in \R^{q}\,$.
	Define $\I_{J}$ according to
	\begin{equation}\label{eq:jump_internal_model}
	\begin{aligned}
	\dot{x}_J = A_J x_J + u_J\,,\quad x^+_J = E_J x_J\,,\quad y_J &= C_J x_J\,,
	\end{aligned}
	\end{equation}
	where $A_J= I_{m_1+n_F} \otimes S\,$, $E_J= I_{m_1+n_F} \otimes J$ and $C_J = I_{m_1+n_F} \otimes C_{J0} = \bbm{C_{J1}' &C_{J2}'}'$ with $C_{J1}\in \R^{m_1\times n_J}\,$, $C_{J2}\in \R^{(m-m_1)\times n_J}\,$ and $n_J = (m_1+n_F)q\,$.
\end{alg}
\subsubsection{The Stabilizer $\K$}\label{subsec:stab}
The design of an output feedback stabilizer for the interconnection, as in \ref{fig:regulator}, of the plant $\PL$ and the internal model $\I_M$ 
is now addressed.\\
In particular, assuming to apply the coordinate change \ref{eq:change_coordinates}, such interconnection has the following form
\begin{subequations}\label{eq:big_system}
\begin{align}
\dot{\xi} &= \hat{A}\xi+ \hat{B}\hat{v}+ \hat{P} w\,,\\
\e &= \hat{C}\xi + \hat{Q} w \,,\\
\xi^+ &= \hat{E}\xi\,,
\end{align}
\end{subequations}
where $\xi = \bbm{z' &x_F' &x_J'}'$,  $\hat{v} = \bbm{u_x' &u_F' &u_J'}'$ and
\begin{equation*}
\begin{aligned}
	\hat{A} &= \bbm{\bar{A} &\bar{B}_2C_F &\bar{B}_1C_{J1} \\ 0 &A_F &0\\  0 &0 &A_J},
	&\hat{C} &= \bbm{\bar{C} &0 &0},\\
	\hat{B} &= \bbm{\bar{B}&0&0\\0 &I_{n_F}&0\\0 &0 &I_{n_J}},
	&\hat{E} &= \bbm{\bar{E} &0 &0 \\ 0 &0 &C_{J2}\\  0 &0 &E_J},\\
	\hat{P} &=\bbm{\bar{P}' &0 &0}',
	&\hat{Q} &=Q\,.
\end{aligned}
\end{equation*}
Since $\PL$ is stabilizable (by \ref{ass:structural_Conditions}), 
it can be shown that also the system in \ref{eq:big_system} is stabilizable,
then the input $\hat{v}$ in \ref{eq:feedback_transformation} can be designed, as described in Appendix \pref{app:stabilizer}, in order to stabilize the system \ref{eq:big_system} with $w \equiv 0\,$.

\section{Example}\label{sec:example}
\begin{figure*}[t]
    \centering
    \includegraphics[width=1.7\columnwidth]{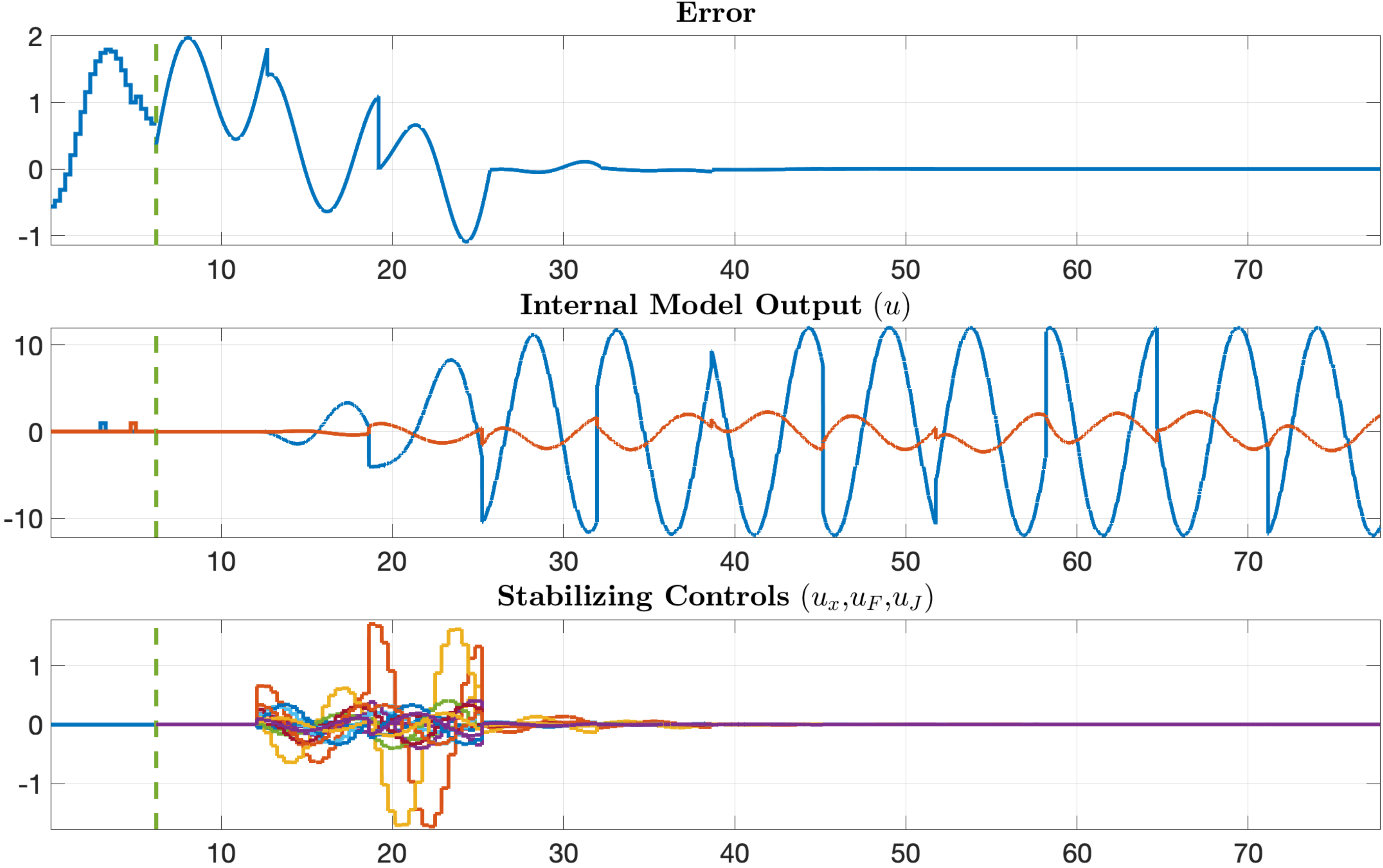}
    \caption{Evolution of the regulated output $\e$ (top) and control inputs, using the regulator scheme in \ref{fig:internal_model}. The vertical dashed line limits the estimation phase.}
    \label{fig:result}
\end{figure*}
Consider the system \ref{eq:linear_plant} described by the matrices
\begin{align*}
{A} &=\bbm{ \text{-}0.505   &0.707    &0\\0.303   &\text{-}0.303        &0\\0.303    &0.707   &\text{-}0.505},
{B} =\bbm{  1.012    &1.012\\0    &1.012\\0    &1.012},\\
{P} &=\bbm{ 0     &0\\0     &0\\0     &0},
{E} =\bbm{ 0.1854    &0.1720    &0.0423\\0.2384   &0.3006    &0.0698\\0.0979    &0.3018    &0.0173},\\
{C} &=\bbm{0  &0    &1.05}\,,
\end{align*}
and the exosystem \ref{eq:exosystem} characterized by
\begin{align*}
{S} =\bbm{      0     &1\\\text{-}1     &0}\,,\quad
{J} =\bbm{      0     &1\\\text{-}1     &0}\,,
\end{align*}
with the initial conditions $x(0)=\bbm{0.559&0.259&0.415}'$, $w(0)=\bbm{1&0}'$ and $\tau_M = 6.5\,$.
Note that, according to the proposed approach, the only information provided for the design of the regulator 
\ref{eq:hybrid_regulator} are $S$, $J$, $\tau_M$ and a nominal description of the plant.
The results obtained by the regulator proposed in \ref{sec:HR} are presented in \ref{fig:result}. Output regulation is achieved as shown in the upper diagram of \ref{fig:result}.
Moreover, in the middle and in the bottom diagrams the evolution of the internal model output and the vanishing action of the stabilizer control are shown, respectively.

\section{Conclusions}\label{sec:conclusions}

This paper proposes the first general solution to the robust hybrid output regulation problem for the class of systems 
proposed in \cite{marconi2010note}, thus removing the restrictive structural assumptions exploited in all previous contributions on the subject.
Future work will deal with the development of alternative solutions, possibly including more efficient adaptation schemes
or some form of optimality, \emph{e.g.} in the LQR sense along the lines in \cite{2014_MEDa,PossieriCDC2016}.

\begin{appendix}
\subsection{Discrete-Time Observer}\label{app:observer}
A straightforward implication of the relations \ref{app:simple_discrete_description1} is that
\begin{align}\label{eq:app_relation}
Y_{[i]}= \tilde{C} \eta_{[i]} + \tilde{D} U_{[i]}\,,
\end{align}
where 
\begin{equation}\label{eq:app_important_matrices}
\begin{aligned}
   \tilde{C}&=\bbm{C'&(CA_D)' &\dots& (CA_D^{n-1})' }'\,,\\
   \tilde{D}&=\bbm{0 &\cdots &0 &0\\
    CB_D &\ddots  &\vdots &\vdots\\\vdots &\ddots  &0&0\\ CA_D^{n-2} B_D &\dots &CB_D&0},
\end{aligned}
\end{equation}
and $Y_{[i]} = [y_{[i]}',\cdots,y_{[i+n-1]}']' $, $U_{[i]} = [u_{[i]}',\dots,u_{[i+n-1]}']'\,$ are a collection of $n$ consecutive measurements of the inputs and the outputs.\\
Consequently, if \ref{eq:linear_plant} is detectable,
a hybrid observer of the following form can be designed
\begin{align*}
    \hat{\eta}_{[h+1]} &=A_D\hat{\eta}_{[h]}+B_Du_{[h]}\,,\quad \text{for $h\in\{0,\dots,n-1\}\,$,}\\
    \hat{\eta}^+_{[n]} &= E\hat{\eta}_{[n]}-L(Y_{[0]}-\tilde{D}U_{[0]}-\tilde{C} \hat{\eta}_{[0]})\,,
\end{align*}
where $L\in \R^{n \times pn}$ is defined in order to obtain that $\Lambda(EA_{D}^n-L\tilde{C}) \subset\C_g\,$.
\subsection{Discrete-Time Stabilizer}\label{app:stabilizer}
The relations \ref{app:simple_discrete_description1} can be exploited to deduce that
\begin{equation*}
    \eta_{[n]}^+= EA_D^n \eta_{[0]}+ \sum_{i=0}^{n-1} EA_D^i B_D u_{[i]}\,,
\end{equation*}
which can be equivalently written as
\begin{align}
\eta_{[n]}^+&= EA_D^n \eta_{[0]}+ E \cdot \mathcal{R}(A_D,B_D) U_{[0]}\,,
\end{align}
with $U_{[0]} = \bbm{u_{[0]}'&\dots&u_{[n-1]}'}'\,$ and $\mathcal{R}(A_D,B_D) := \bbm{B_D &A_DB_D &\cdots &(A_D)^{n-1}B_D}\,$.\\
Consequently, if a condition like \ref{eq:PBH_test_b} holds for the system \ref{eq:linear_plant},
a control $U_{[0]}:= K \hat{\eta}_{[0]}$, with $\hat{\eta}_{[0]}$ as $\eta_{[0]}$ whenever its measurements are available or otherwise as an estimation of $\eta_{[0]}$ achieved with an observer like the one described in Appendix \pref{app:observer}, can be defined in order to obtain that $\Lambda(EA_D^n+E \cdot \mathcal{R}(A_D,B_D)K) \subset\C_g\,$.
\end{appendix}




\bibliographystyle{IEEEtran}

\bibliography{our_bib}

\begin{thebibliography}{10}
\providecommand{\url}[1]{#1}
\csname url@samestyle\endcsname
\providecommand{\newblock}{\relax}
\providecommand{\bibinfo}[2]{#2}
\providecommand{\BIBentrySTDinterwordspacing}{\spaceskip=0pt\relax}
\providecommand{\BIBentryALTinterwordstretchfactor}{4}
\providecommand{\BIBentryALTinterwordspacing}{\spaceskip=\fontdimen2\font plus
\BIBentryALTinterwordstretchfactor\fontdimen3\font minus
  \fontdimen4\font\relax}
\providecommand{\BIBforeignlanguage}[2]{{%
\expandafter\ifx\csname l@#1\endcsname\relax
\typeout{** WARNING: IEEEtran.bst: No hyphenation pattern has been}%
\typeout{** loaded for the language `#1'. Using the pattern for}%
\typeout{** the default language instead.}%
\else
\language=\csname l@#1\endcsname
\fi
#2}}
\providecommand{\BIBdecl}{\relax}
\BIBdecl

\bibitem{francis1976internal}
B.~Francis and W.~Wonham, ``The internal model principle of control theory,''
  \emph{Automatica}, vol.~12, no.~5, pp. 457--465, 1976.

\bibitem{davison1976robust}
E.~Davison, ``The robust control of a servomechanism problem for linear
  time-invariant multivariable systems,'' \emph{IEEE Trans. Aut. Cont.},
  vol.~21, no.~1, pp. 25--34, 1976.

\bibitem{Wonham1985}
W.~Wonham, \emph{{Linear Multivariable Control. A Geometric Approach}},
  3rd~ed., ser. Applications of Mathematics.\hskip 1em plus 0.5em minus
  0.4em\relax New York, NY: Springer Verlag, 1985, vol.~10.

\bibitem{byrnes1997output}
C.~I. Byrnes, F.~Delli~Priscoli, and A.~Isidori, \emph{Output regulation of
  uncertain nonlinear systems}.\hskip 1em plus 0.5em minus 0.4em\relax
  Birkh{\"a}user, 1997.

\bibitem{huang}
J.~Huang, \emph{Nonlinear output regulation: Theory and applications}.\hskip
  1em plus 0.5em minus 0.4em\relax Society for Industrial Mathematics, 2004.

\bibitem{pavlov2005uniform}
A.~V. Pavlov, N.~van~de Wouw, and H.~Nijmeijer, \emph{Uniform output regulation
  of nonlinear systems: a convergent dynamics approach}.\hskip 1em plus 0.5em
  minus 0.4em\relax Birkh{\"a}user Boston, 2005.

\bibitem{serrani_adaptiveIM}
A.~Serrani, A.~Isidori, and L.~Marconi, ``Semi-global nonlinear output
  regulation with adaptive internal model,'' \emph{IEEE Trans. Aut. Cont.},
  vol.~46, no.~8, pp. 1178--1194, 2001.

\bibitem{de2018data}
G.~de~Carolis, S.~Galeani, and M.~Sassano, ``Data-driven, robust output
  regulation in finite time for {LTI} systems,'' \emph{Int. J. Robust and
  Nonlinear Control}, vol.~28, no.~18, pp. 5997--6015, 2018.

\bibitem{2016_IJACSP}
D.~Carnevale, S.~Galeani, M.~Sassano, and A.~Astolfi, ``Robust hybrid
  estimation and rejection of multi-frequency signals,'' \emph{International
  Journal of Adaptive Control and Signal Processing}, vol.~30, no.~12, pp.
  1649--1673, 2016, acs.2679.

\bibitem{Bodson1994}
M.~Bodson, A.~Sacks, and P.~Khosla, ``{Harmonic generation in adaptive
  feedforward cancellation schemes},'' \emph{IEEE Trans. Aut. Cont.}, vol.~39,
  no.~9, pp. 1939--1944, 1994.

\bibitem{Messineo2009a}
S.~Messineo and A.~Serrani, ``{Adaptive feedforward disturbance rejection in
  nonlinear systems},'' \emph{Systems \& Control Letters}, vol.~58, no.~8, pp.
  576--583, 2009.

\bibitem{serrani2006rejection}
A.~Serrani, ``Rejection of harmonic disturbances at the controller input via
  hybrid adaptive external models,'' \emph{Automatica}, vol.~42, no.~11, pp.
  1977--1985, 2006.

\bibitem{2017_IFACWCa}
D.~Carnevale, S.~Galeani, M.~Sassano, and A.~Serrani, ``External models for
  output regulation based on moment estimation from input-output data,'' in
  \emph{IFAC World Congress}, Jul. 2017, pp. 8043--8048.

\bibitem{2017_MEDb}
G.~de~Carolis, S.~Galeani, and M.~Sassano, ``Data driven, robust output
  regulation via external models,'' in \emph{Mediterranean Control Conf.},
  2017, pp. 1263--1268.

\bibitem{Saberi}
A.~Saberi, A.~Stoorvogel, and P.~Sannuti, \emph{Control of linear systems with
  regulation and input constraints}.\hskip 1em plus 0.5em minus 0.4em\relax
  Springer, 2000.

\bibitem{galeani2015input}
S.~Galeani, A.~Serrani, G.~Varano, and L.~Zaccarian, ``On input
  allocation-based regulation for linear over-actuated systems,''
  \emph{Automatica}, vol.~52, pp. 346--354, 2015.

\bibitem{CDC_2011b}
S.~Galeani, A.~Serrani, G.~Varano, and L.~Zaccarian, ``On linear over-actuated
  regulation using input allocation,'' in \emph{Conf. on Decision and Control},
  2011, pp. 4471--4476.

\bibitem{2018_ECCb}
M.~Sassano and S.~Galeani, ``Output regulation for redundant plants via
  orthogonal moments,'' in \emph{European Control Conf.}, 2018, pp. 1945 --
  1950.

\bibitem{2018_CDC}
S.~Galeani and M.~Sassano, ``Data-driven dynamic control allocation for
  uncertain redundant plants,'' in \emph{Conf. on Decision and Control}, Dec.
  2018, pp. 5494 -- 5499.

\bibitem{GMP_IJC_2008}
S.~Galeani, L.~Menini, A.~Potini, and A.~Tornambe, ``Trajectory tracking for a
  particle in elliptical billiards,'' \emph{International Journal of Control},
  vol.~81, no.~2, pp. 189--213, 2008.

\bibitem{morarescu2010trajectory}
I.~Morarescu and B.~Brogliato, ``Trajectory tracking control of multiconstraint
  complementarity lagrangian systems,'' \emph{IEEE Trans. Aut. Cont.}, vol.~55,
  no.~6, pp. 1300--1313, 2010.

\bibitem{Forni}
F.~{Forni}, A.~R. {Teel}, and L.~{Zaccarian}, ``Follow the bouncing ball:
  Global results on tracking and state estimation with impacts,'' \emph{IEEE
  Trans. Aut. Cont.}, vol.~58, no.~6, pp. 1470--1485, 2013.

\bibitem{GMP_TAC12}
S.~Galeani, L.~Menini, and A.~Potini, ``Robust trajectory tracking for a class
  of hybrid systems: an internal model principle approach,'' \emph{IEEE Trans.
  Aut. Cont.}, vol.~57, no.~2, pp. 344--359, 2012.

\bibitem{marconi2010note}
L.~Marconi and A.~R. Teel, ``A note about hybrid linear regulation,'' in
  \emph{Conf. on Decision and Control}.\hskip 1em plus 0.5em minus 0.4em\relax
  IEEE, 2010, pp. 1540--1545.

\bibitem{marconi2013internal}
L.~Marconi and A.~R. Teel, ``Internal model principle for linear systems with
  periodic state jumps,'' \emph{IEEE Trans. Aut. Cont.}, vol.~58, no.~11, pp.
  2788--2802, 2013.

\bibitem{ZATTONI2017322}
E.~Zattoni, A.~M. Perdon, and G.~Conte, ``Output regulation by error dynamic
  feedback in hybrid systems with periodic state jumps,'' \emph{Automatica},
  vol.~81, pp. 322 -- 334, 2017.

\bibitem{CDC_2012a}
D.~Carnevale, S.~Galeani, and L.~Menini, ``Output regulation for a class of
  linear hybrid systems. {P}art 1: trajectory generation,'' in \emph{Conf. on
  Decision and Control}, 2012.

\bibitem{CDC_2012b}
D.~Carnevale, S.~Galeani, and L.~Menini, ``Output regulation for a class of
  linear hybrid systems. {P}art 2: stabilization,'' in \emph{Conf. on Decision
  and Control}, 2012.

\bibitem{carnevale2016hybrid}
D.~Carnevale, S.~Galeani, L.~Menini, and M.~Sassano, ``Hybrid output regulation
  for linear systems with periodic jumps: Solvability conditions, structural
  implications and semi-classical solutions,'' \emph{IEEE Trans. Aut. Cont.},
  vol.~61, no.~9, pp. 2416--2431, 2016.

\bibitem{2013_MEDc}
D.~Carnevale, S.~Galeani, and L.~Menini, ``A case study for hybrid regulation:
  output tracking for a spinning and bouncing disk,'' in \emph{Mediterranean
  Control Conf.}, Jun. 2013, pp. 858--867.

\bibitem{carnevale2017robust}
D.~Carnevale, S.~Galeani, L.~Menini, and M.~Sassano, ``Robust hybrid output
  regulation for linear systems with periodic jumps: Semiclassical internal
  model design,'' \emph{IEEE Trans. Aut. Cont.}, vol.~62, no.~12, pp.
  6649--6656, 2017.

\bibitem{2017_CDCc}
G.~de~Carolis, S.~Galeani, and M.~Sassano, ``Data-driven deadbeat control with
  application to output regulation,'' in \emph{Conf. on Decision and Control},
  2017, pp. 6271--6276.

\bibitem{trentelman2002control}
H.~L. Trentelman, A.~A. Stoorvogel, and M.~Hautus, \emph{Control theory for
  linear systems}, 2002.

\bibitem{2014_MEDa}
D.~Carnevale, S.~Galeani, and M.~Sassano, ``A linear quadratic approach to
  linear time invariant stabilization for a class of hybrid systems,'' in
  \emph{Mediterranean Control Conf.}, 2014.

\bibitem{PossieriCDC2016}
C.~{Possieri} and A.~R. {Teel}, ``{LQ} optimal control for a class of hybrid
  systems,'' in \emph{Conf. on Decision and Control}, 2016, pp. 604--609.

\end{thebibliography}

\end{document}